# Using Blippar Augmented Reality Browser in the Practical Training of Mechanical Engineers

Andrii Striuk[1][0000-0001-9240-1976], Maryna Rassovytska[1][0000-0003-4973-0082]
and Svitlana Shokaliuk[2][0000-0003-3774-1729]

[1] State Institution of Higher Education «Kryvyi Rih National University»
11, Vitali Matusevich st., Kryvyi Rih, 50027 Ukraine
`andrey.n.stryuk@gmail.com, rassovitskayamarina@mail.ru`
[2] Kryvyi Rih State Pedagogical University
54, Gagarina Ave., Kryvyi Rih, 50086 Ukraine
`shokalyuk15@gmail.com`

**Abstract.** The purpose of the study is to justify the expediency of using the Blippar augmented reality browser for professional and practical training of future mechanical engineers. Tasks of the research: to analyze the expediency of using augmented reality tools in the professional training of bachelors of applied mechanics; to carry out the selection of augmented reality tools, which is expedient to use in the training of future engineer mechanics; to develop educational materials using the chosen augmented reality tools. The object of the study is the professional training of future mechanical engineers. The subject of the study is the use of the augmented reality tools in the professional training of bachelors of applied mechanics. The paper analyzes the relevance and expediency of the use of the augmented reality tools in the professional training of future mechanical engineers. It is determined that the augmented reality tools will promote the development of ICT competence and graphic competence of bachelors of applied mechanics The model of the use of the augmented reality tools in the training of future mechanical engineers is proposed. As the main tool, the Blippar browser and Blippbuilder's cloud-based script development tool are chosen. An example of the creation of markers and scenes of augmented reality using the selected tools is given. The advantages and disadvantages of used tools are indicated. The proposed learning tools and methods can be applied to vocational and practical training of mechanical engineers.

## 1 Introduction and Problem Statement

Mechanical engineering today is the basis of the development of high technologies: its integration and convergence with electrical engineering, automation and computer-integrated technologies and software engineering have created a new direction in the training of the most sought-after specialists – mechatronics. The need to ensure the competitiveness of domestic mechanical engineers in this area at the domestic, Euro-pean and world market of high-intellectual work requires a rethinking and deepening of the ICT component of bachelors in applied mechanics based on the trends of ICT development in the fields of engineering, science and education: from the geographical mobility of workers to the mobility of professional and professionally oriented ICT tools [1].

Higher education institutions facilitate the training of engineers, implementation of innovations in education and training, in particular the strengthening of practical orientation, the introduction of a problem-based learning system, and the widespread use of progressive ICT. The realization of this direction requires the purposeful formation of ICT competence of bachelors in applied mechanics: both traditional (mastering new technologies, forming competencies in the selection, analysis, systematization and evaluation of professionally directed information) and innovative (developing and adapting ICT tools to support professional activities, management of professional self-development, creation of new technologies) [2].

The leading tools of forming the ICT competencies of future engineers – mechanics are automation tools for designing and documenting new parts, mechanisms and machines – specialized CAD systems [3]. The use of CAD systems is also due to the formation of graphic competence, the development of which was studied by O. M. Jedzhula [4]. The researcher emphasizes that the professional competence of an engineer is largely determined by the ability to learn about a technical object or its principles of design documentation, to capture information in graphical form, to use a graphic image for the purpose of communication; to make an expedient decision in the conditions of modern technogenic society, using graphic means and methods, computer graphic products [4]. Graphic training of students requires taking into account perspective directions in the professional activity of modern engineer (computer engineering, design, ergonomics), deployment of effective intellectual communications, conceptual (concentrically integral) development of graphical knowledge – gradual transition from the stage of visual-graphic modeling to computer modeling of 3D-objects (computer graphics) and four-dimensional simulation of 3D-objects (computer animation), hierarchical structure of graphic activity, polyfunctionality of graphic images.

Modern CAD systems provide the opportunity to work both with 2D-drawings, and with 3D-models. But their use requires significant computing resources that are not always available during classroom or independent work. That is why, despite the rapid development of ICT, paper carriers have not lost their relevance for design and engineering documentation. At the same time, paper drawings lack interactivity. It is possible to compensate for these disadvantages using the technology of augmented reality (AR) [5], which enable the possibility of combining real-world images with virtual objects, for example, 3D-models [6]. An important feature of using the AR tools is the absence of the need to use resource-intensive software for building 3D-images. To display the AR objects is enough mobile device with the installed program – an augmented reality browser, e. g. Blippar [7].

## 2 Methods of using Blippar as a tool of training future engineer mechanics

Blippar is one of the most commonly used AR browsers. The main part of Blippar is a subsystem of pattern recognition, through which students looking for markers. Markers can be real world objects or images. After recognizing the marker, the program adds virtual objects (blipps) to the real image, which can be images, 3D- models, animations, interactive dynamic scenes, links, etc. For the training of future engineers, the model of Blippar using, shown in Fig. 1. The main AR object is a 3D-model illustrating a flat image on a marker. A student can manage a model, its scale, turning, etc. From the model review student can go to additional materials: a text description, a gallery of images, a demonstration video, etc. Markers are created using the Blippbuilder Cloud Tool [8]. Using Blippbuilder for educational purposes is free. Teachers have access to all functionalities without limitations.

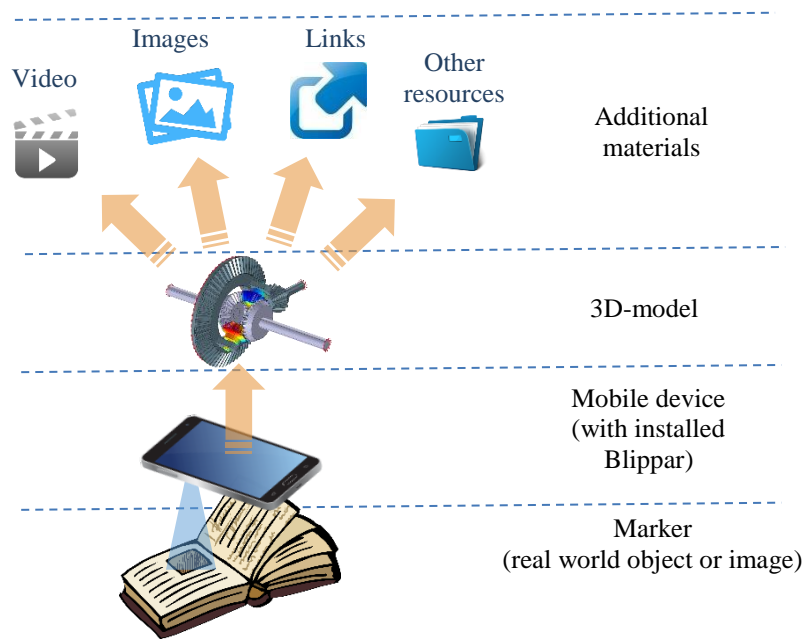

**Fig. 1.** The model of using Blippar in training future mechanical engineers

But for access to learning blipps students need to enter a special code generated by the program. Blippar developers recommend that you save all created content in one "campaign", as it allows you to publish a single, unified "campaign code" that students can use to unlock all of your blipps. Using this approach, students only need to enter the code once on their device, and never have to change it. When an account is created for school or college, the first registered user becomes the administrator of this group. As an administrator, you can freely add additional members to the group, both the faculty and the students. Students should also download the Blippar application on their mobile device to view the content created by Blippbuilder. The work of the teacher on creating a new AR object begins with the choice of a marker. A marker can be any image: drawings, drawing, photography. Photos and drawings are mostly better recognized by Blippar, so it's better to use them.

To demonstrate the use of Blippar in the training of future mechanical engineers, we have chosen the image of the toothed wheel (Fig. 2).

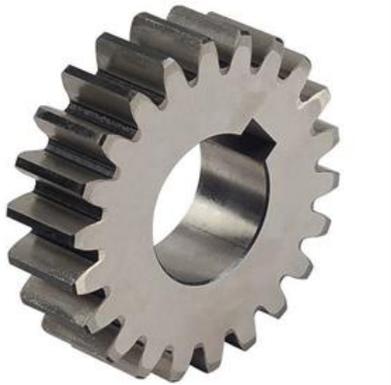

**Fig. 2.** Marker Images

To create a new blipp, go to the https://hub.blippar.com/ link, sign in and authorize. After that you will have access to a personal office where you can create and edit new blipps (Fig. 3).

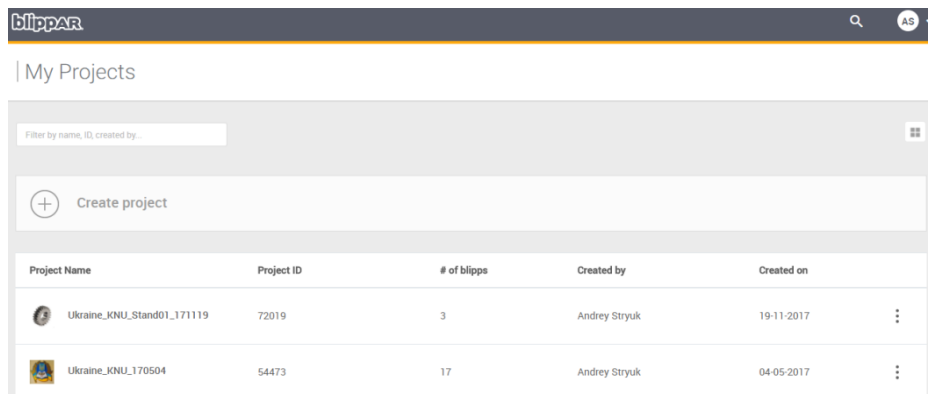

**Fig. 3.** User account

Blipps are created and edited within a particular project, so first of all you need to choose an existing one or create a new project. Then you can create new blipps (Fig. 4).

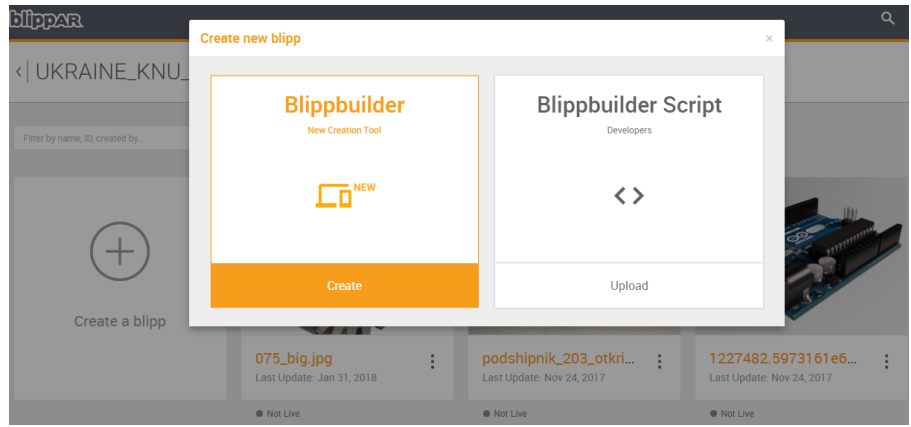

**Fig. 4.** Creating a new blipp

Creating a blipp begins with uploading a marker image. After uploading the marker, you will be taken to the Blippbuilder homepage. You will see your marker in the center of the screen. This is your scene, which you can now add a variety of virtual objects: images, video, audio or 3D models (Fig. 5).

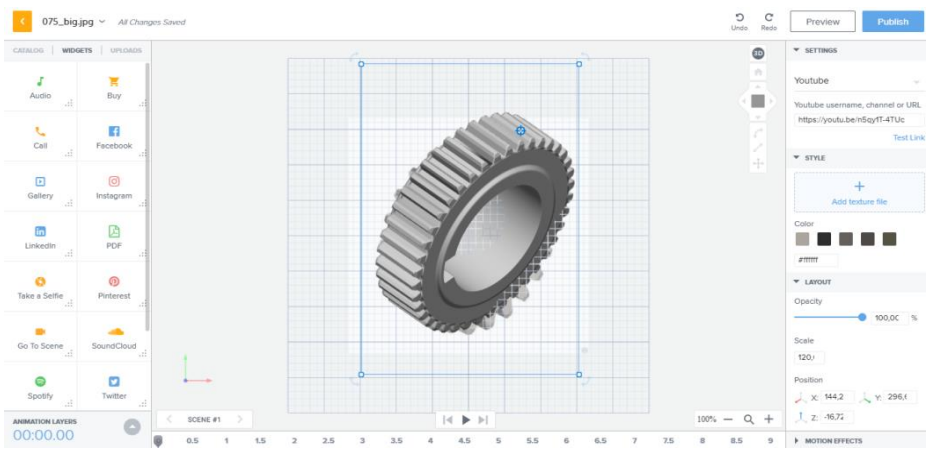

**Fig. 5.** Scene of creating a new blipp

After you put all your resources on the stage, you can resize, move and rotate them with the mouse. For more precise control, you can use the transformation controls in the right panel (Fig. 5).

While working on a scene, you have the ability to check marker recognition and scene display on your mobile device. Click the "Preview" button in the upper right corner of the screen. A popup window appears displaying the status of your blipp download. When the download is completed, follow the on-screen instructions to test the blipp (Fig. 6).

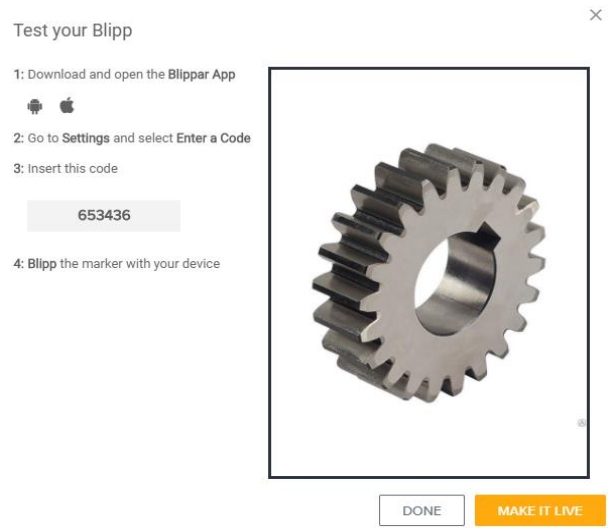

**Fig. 6.** Blipp test

For educational materials, the blipp testing code is used as a public access code. It must be reported to the students or printed next to the marker. After recognizing the marker, Blippar will display the downloaded 3D model by mixing it with the image from the camera device (Fig. 7).

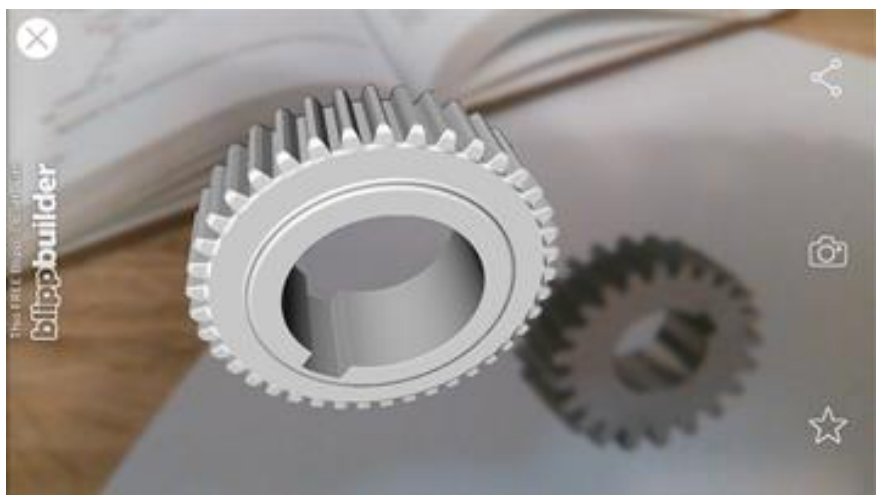

**Fig. 7.** Display of three-dimensional model

In our case, additional materials are attached to the model, in particular a videoclip showing the work of the toothed wheel (Fig. 8).

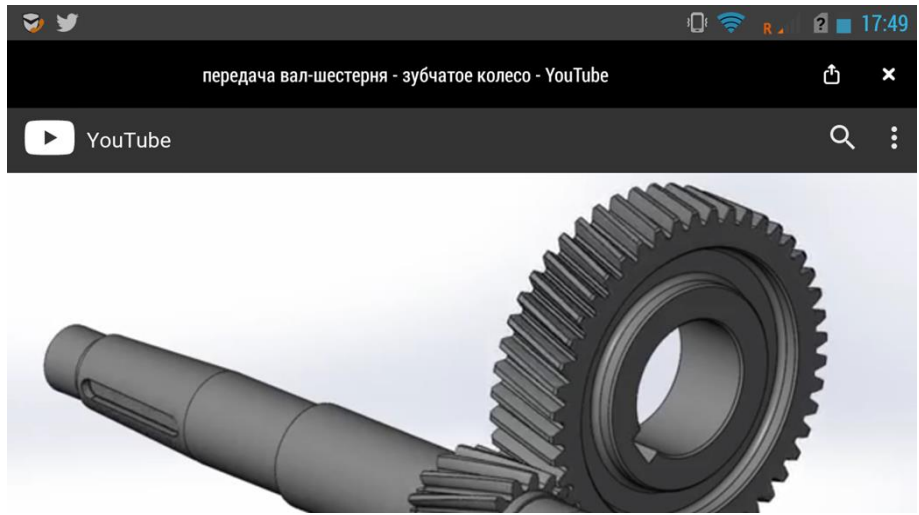

**Fig. 8.** Demo Video

## 3 Conclusion

1. The expediency and feasibility of using the technology of augmented reality in the training of future mechanical engineers was shown in this article, proposes a model for using the Blippar and examines the process of creating scenes of augmented reality using the cloud service Blippbuilder.
2. The software that has been considered has a powerful enough functionality to develop scenarios and scenarios of augmented reality using multimedia materials, and in particular three-dimensional models. At the same time, certain drawbacks of the programs under consideration were identified: (a) the need for student access code; (b) limited number of formats for the representation of 3D-models; (c) slow work of programs during visualization of complex 3D-models.
3. Proceeding from the trends of ICT development, in the near future we should expect significant improvement of both software and hardware, through which the technology of augmented reality is realized. This will promote both wider use of this technology in various spheres of human activity, as well as overcoming the existing shortcomings.